\documentclass[final,authoryear,12pt]{elsarticle}

\usepackage{graphicx}
\usepackage{amsthm}
\usepackage[fleqn]{amsmath}
\usepackage{subfigure}
\usepackage{epstopdf}
\usepackage{multirow}
\usepackage{booktabs} % for professional tables
\usepackage{slashbox}
\usepackage[draft,bookmarks]{hyperref}
\hypersetup{colorlinks=true,linkcolor=black,urlcolor=black,citecolor=black,bookmarks=true}
\usepackage{geometry}                		% See geometry.pdf to learn the layout options. There are lots.
\usepackage{graphicx}				% Use pdf, png, jpg, or epsǠwith pdflatex; use eps in DVI mode
\usepackage{lscape}								% TeX will automatically convert eps --> pdf in pdflatex		
\usepackage{amssymb}
\usepackage[usenames]{xcolor}
\usepackage{soul} % for highlighting text
\usepackage[utf8]{inputenc}

\biboptions{comma,round}

\journal{Working Paper}

\begin{document}

\begin{frontmatter}

\title{Can we still benefit from international diversification? The case of the Czech and German stock markets\tnoteref{label1}}

\author[ies,utia]{Krenar Avdulaj}
 \author[ies,utia]{Jozef Barunik\corref{cor2}} \ead{barunik@utia.cas.cz}
 \address[ies]{Institute of Information Theory and Automation, Academy of Sciences of the Czech Republic, Czech Republic}
\address[utia]{Institute of Economic Studies, Charles University, Prague, Czech Republic}

\tnotetext[label1]{We are indebted to two anonymous referees for their helpful comments. Jozef Barunik gratefully acknowledges the support of the Czech Science Foundation project No. P402/12/G097 DYME - ``Dynamic Models in Economics". Krenar Avdulaj gratefully acknowledges the support of the Grant Agency of the Charles University under the 852013 project.}

\begin{abstract}

One of the findings of the recent literature is that the 2008 financial crisis caused reduction in international diversification benefits. To fully understand the possible potential from diversification, we build an empirical model which combines generalised autoregressive score copula functions with high frequency data, and allows us to capture and forecast the conditional time-varying joint distribution of stock returns. Using this novel methodology and fresh data covering five years after the crisis, we compute the conditional diversification benefits to answer the question, whether it is still interesting for an international investor to diversify. As diversification tools, we consider the Czech PX and the German DAX broad stock indices, and we find that the diversification benefits strongly vary over the 2008--2013 crisis years.
\end{abstract}

\end{frontmatter}

\section{Introduction}

A proper quantification of the joint distribution allowing for the time-varying dependence between assets is critical for asset pricing, portfolio allocation and risk reduction. For a number of years, finance literature has been studying the risk reduction benefit from international diversification. After the recent 2008 financial crisis, many researchers have documented possible reduction of these benefits due to rising dependence between markets. The literature concentrating on the Central European Markets has been limited though, as it is widely believed that after the enlargement of the European Union, these markets became integrated with very limited opportunities for diversification. 

In this paper, we revisit this line of research, and study the possible benefits from diversification between the Czech PX and the German DAX stock market indices using data covering the five years crisis period. While it is reasonable to believe that the Czech and German stock markets show large degree of dependence due to integration of the Czech Republic into the euro area as well as large dependence of the Czech economy on the German one, we aim to study, whether the German and Czech stock indices can be considered for the reduction of risk of an international investor.

A number of researchers have addressed the issue of Central and Eastern European (CEE) markets integration with the euro area. \cite{voronkova2004equity} documents increasing stock integration between Central European (CE) markets and their mature counterparts in Europe, and finds lower diversification opportunities at the aggregate stock index level. \cite{syriopoulos2004international,syriopoulos2006risk} finds long-run cointegrating relationship and hence limited diversification opportunities between CEE markets and Germany. \cite{aslanidis2011there} confirm these findings using more recent data. \cite{Egert2007} analyze the intraday interdependence of Western European and Central and Eastern European markets using wide range of econometric techniques and find evidence of only short-term relationships among the CEE and Western European stock markets.

In more recent study, \cite{Egert2011} analyze the comovements of three developed and three emerging markets using DCC-GARCH model on high frequency data. They detect small correlation between the developed and emerging markets. This finding is important for investors, as it allows them to diversify their portfolios by investing in the emerging markets. However, as the authors stress, this diversification opportunity may not be available more recently due to economic integration with Western Europe. A study by \cite{Hanousek2011} use high frequency data to study the foreign macroeconomic announcements and spillover effects on emerging CEE stock markets during 2004-2007. Among other findings it is of interest that Frankfurt stock market dominates the spillovers over the three emerging markets, while the reaction to the New York market is smaller.  \cite{Syllignakis2011} study the time-varying conditional correlations among the US, Germany, Russia, and CEE markets. The authors employ the DCC-GARCH model for correlations and use weekly data spanning from 1997-2009 and they find significant increase of correlation between the US and German markets and the CEE markets, especially during 2007-2009 financial crisis. 

In a recent work, \cite{Horvath2013} analyze the comovements among Western Europe vis-\`a-vis Central and South Eastern Europe (SEE). The analysis is carried on daily data for the period 2006-2011 using bivariate BEKK-GARCH model. Authors find the higher integration among CE and lower integration, with almost zero correlations, among SEE countries. In another study, \cite{Gjika2013} employ asymmetric DCC-GARCH to study the comovements in CE markets. Using daily data spanning from 2001-2011 authors find increase in correlations after these countries joined the European Union, whereas asymmetric correlation effects were found only for Hungary (BUX) and Poland (WIG) pair. In addition, positive relation among conditional correlations and conditional variances suggesting lower diversifications in turbulent times is confirmed.

A common feature of these studies is that they use cointegration, or multivariate GARCH to study the dependence, and with some exceptions, they use the data before the 2008 financial crisis. We contribute to the literature by using a very different approach proposed recently by \cite{Avdulaj2013}, allowing us to model the time-varying joint distribution of stock market returns. Using recently proposed time-varying copula methodology and utilizing high frequency data, we build an empirical model which allows us to study the time-varying benefits from diversification. In addition, we contribute to the understanding of the relationship using recent data covering crisis years. Using the fresh data, and state of the art methodologies, we revisit the literature and uncover significant time-varying nature of the benefits from diversification between the PX and DAX markets. This finding is particularly interesting as previous literature generally reports decreasing potential for diversification.

The work is organized as follows. The second section introduces our empirical model composing from the realized GARCH and generalised autoregressive score time-varying copulas. The third section introduces the data we use, while fourth section discusses the in-sample and out-of-sample fits of all model specifications, and chooses the one which best describes the data. Finally, the fifth section tests the economic implications of our empirical model. We first evaluate the quantile forecasts, which are central to risk management, and then study the time-varying diversification benefits implied by our model. The last section concludes. 

\section{Dynamic copula realized GARCH modeling framework}

We introduce the empirical model used for describing the dependence between the German DAX and Czech PX stock indices. Our modeling strategy utilizes high frequency data to capture the dependence in the margins and recently proposed dynamic copulas to model the dynamic dependence. Final model is thus able to describe the conditional time-varying joint distribution of returns.

The methodology is based on the \citeauthor{Sklar1959}'s (1959) theorem extended to conditional distributions by \cite{Patton2006}. The extended \citeauthor{Sklar1959}'s theorem allows to decompose a conditional joint distribution into marginal distributions and a time-varying copula. Consider the bivariate stochastic process $\{\mathbf{X}_t\}_{t=1}^T$ with $\mathbf{X_t} = (X_{1t},X_{2t})'$, which has a conditional joint distribution $\mathbf{F_t}$ and conditional marginal distributions $F_{1t}$ and $F_{2t}$. Then
\begin{equation}
\mathbf{X_t}|\mathcal{F}_{t-1} \sim \mathbf{F_t} = \mathbf{C_t}\left(F_{1t},F_{2t}\right),
\end{equation}
where $\mathbf{C_t}$ is the time-varying conditional copula of $\mathbf{X_t}$ containing all information about the dependence between $X_{1t}$ and $X_{2t}$, and $\mathcal{F}_{t-1}$ available information set, usually $\mathcal{F}_{t} = \sigma(\mathbf{X}_t,\mathbf{X}_{t-1},\ldots)$. Due to \citeauthor{Sklar1959}'s theorem, we are able to construct a dynamic joint distribution $\mathbf{F_t}$ by linking together any two marginal distributions $F_{1t}$ and $F_{2t}$ with any copula function.\footnote{ Note that the information set for the margins and the copula conditional density is the same.}
 Theoretically, there is limitless number of valid joint distribution functions that can be created by combining different copulas with different margins, making this approach very flexible. 
 
\subsection{Time-varying conditional marginal distribution with realized measures}

The first step in building an empirical model based on copulas is to model the margins. Since the largest part of the dependence in financial time series is in their variance, majority of researchers use the generalized autoregressive conditional heteroscedasticity (GARCH) approach of \cite{Boll86} in this step. 

We use the latest advances in the literature which improve volatility modelling by adding the realized volatility measure to the GARCH model. This approach utilizes high frequency data to help in explaining the latent volatility. Compared with standard GARCH(1,1) model where the conditional variance of $i$-th asset, $h_{it}=var(X_{it}|\mathcal{F}_{t-1})$ is dependent on its past values $h_{it-1}$ and past values of $X_{it-1}^2$, \cite{Hansen2012} propose to utilize realized volatility measure and make $h_{it}$ dependent on the realized variance as well. In this work, we restrict ourselves to the simple log-linear specification of the so-called realized GARCH(1,1). For the general framework of realized GARCH($p$,$q$) models we suggest to consult \cite{Hansen2012}. While it is important to model conditional time-varying mean $E(X_{it}|\mathcal{F}_{t-1})$, we also include the standard autoregressive (AR) term into the final modeling strategy. As we will find later, autoregressive term of order no larger than two is appropriate for the DAX and PX return series, thus we restrict ourselves to specifying AR(2) with log-linear RealGARCH(1,1) model as in \cite{Hansen2012}  
\begin{align}
\label{realizedgarch}
X_{it}&=\mu_i + \alpha_1X_{it-1} + \alpha_2X_{it-2} +  \sqrt{h_{it}}z_{it}, \hspace{3cm} \mbox{for  }  i=1,2 \\
\log h_{it}&=\omega_i + \beta_i \log h_{it-1} + \gamma_i \log RV_{it-1}, \\
\log RV_{it}&=\psi_i + \phi_i \log h_{it} + \tau_i(z_{it}) + u_{it},
\end{align}

where $\mu_i$ is the constant mean, $h_{it}$ conditional variance, which is latent, $RV_{it}$ realized volatility measured from high frequency data, $u_{it} \sim N(0,\sigma^2_{iu})$, and $\tau_i(z_{it})=\tau_{i1} z_{it} + \tau_{i2} (z_{it}^2-1)$ leverage function. For the $RV_{it}$, we use the high frequency data and compute it as a sum of squared intraday returns \citep{abdl2003,barndorff2004b}. Innovations $z_{it}$ are modelled by the flexible skewed-\emph{t} distribution of \cite{Hansen1994}. This distribution has two shape parameters, a skewness parameter $\lambda\in(-1,1)$ controlling the degree of asymmetry, and a degree of freedom parameter $\nu\in(2,\infty]$ controlling the thickness of tails. When $\lambda=0$, the distribution reduces to the standard Student's $t$ distribution, and when $\nu \rightarrow\infty$, it becomes skewed Normal distribution, while for  $\nu \rightarrow\infty$ and $\lambda=0$, it becomes $N(0,1)$. 

Thus after the time varying dependence in mean and volatility is modeled, we are left with innovations
\begin{align}
 \hat{z}_{it}&=\frac{X_{it}-\hat{\mu}_i-\hat{\alpha}_1X_{it-1} + \hat{\alpha}_2X_{it-2}}{ \sqrt{\hat{h}_{it}}} \\
 \hat{z}_{it}|\mathcal{F}_{t-1}&  \sim F_i(0,1), \hspace{3cm} \hbox{for } i=1,2.
\end{align}
which have a constant conditional distribution with zero mean and unit variance.  Then the conditional copula of $\mathbf{X_t}|\mathcal{F}_{t-1}$ is equal to the conditional distribution of $\mathbf{U_t}|\mathcal{F}_{t-1}$:\footnote{Since the probability integral transform is invertible, the copula function describes also the dependence of the returns $\mathbf{X_t}|\mathcal{F}_{t-1}$.}
\begin{equation}
\mathbf{U_t}|\mathcal{F}_{t-1} \sim \mathbf{C_t(\gamma_0)},
\end{equation}
with $\gamma$ being copula parameters, and $\mathbf{U_t}=[U_{1t},U_{2t}]'$ conditional probability integral transform 
\begin{equation}
U_{it} = F_{i} \left(\hat{z}_{it};\phi_{i,0} \right), \hspace{3cm} \hbox{for } i=1,2.
\end{equation}

\subsection{Dynamic copulas: A ``GAS" dynamics in parameters}

The notion of time-varying copula models was initially introduced by \cite{Patton2006}. In further literature, \cite{Lee2009} develop a model where the multivariate GARCH is extended by copula functions to capture the remaining dependence. Recently, \cite{hafner2012,manner2011} propose a stochastic copula models, which allow parameters to evolve as a latent time series. Another possibility is offered by ARCH-type models for volatility \citep{Engle2002} and related models for copulas \citep{Patton2006,Creal2013}, which allow the parameters to be some function of lagged observables. An advantage of the second approach is that it avoids the need to ``integrate out" the innovation terms driving the latent time series processes. 

When working with time-varying copula models the driving dynamics of the model is of crucial importance. For our empirical model, we therefore adopt the generalized autoregressive score (GAS) model of \cite{Creal2013}, which specifies the time-varying copula parameter ($\delta_t$) as a function of the lagged copula parameter and a forcing variable that is related to the standardized score of the copula log-likelihood\footnote{ \cite{harvey2013,harvey2012} propose a similar method for modelling time-varying parameters, which they call a dynamic conditional score model.}. This type of dynamics reduces the one-step-ahead prediction error at current observation given the current parameter values of the copula function. Consider a copula with time-varying parameters:
\begin{equation}
\mathbf{U_t}|\mathcal{F}_{t-1} \sim \mathbf{C_t(\delta_t(\gamma))}.
\end{equation}

Often, a copula parameter is required to fall within a specific range e.g. correlation for Normal or student's $t$ copula is required to fall in between values of -1 and 1. To ensure this, \cite{Creal2013} suggest to transform copula parameter by an increasing invertible function $h(\cdot)$ (e.g., logarithmic, logistic, etc.) to the parameter
\begin{equation}
\kappa_t=h(\delta_t) \Longleftrightarrow  \delta_t= h^{-1}(\kappa_t) \label{eq:invertft}
\end{equation}
For a copula with transformed time-varying parameter $\kappa_t$, a GAS(1,1) model is specified as
\begin{align}
 \kappa_{t}&= w + \beta \kappa_{t-1}+ \alpha I^{-1/2}_t \mathbf{s}_{t-1} \\
\mathbf{s}_{t-1} &\equiv\displaystyle \frac{ \partial \log \mathbf{c} (\mathbf{u_{t-1}};\delta_{t-1})}{\partial \delta_{t-1}} \\
I_t &\equiv E_{t-1}[\mathbf{s}_{t-1}\mathbf{s}_{t-1}']=I(\delta_t).
\end{align}
While this specification for the time-varying parameters is arbitrary, \cite{Creal2013} motivate it in a way that the model nests a variety of popular approaches from conditional variance models to trade durations and counts models. Also, the recursion is similar to numerical optimisation algorithms such as the Gauss-Newton algorithm.

Having specified the model, the last step is to choose the copula function used in the application. The time-varying copulas we use in this work are the rotated Gumbel, Normal and Student's $t$. In addition we use constant copulas for comparison. To save the space, we do not provide functional forms of copula functions used in this work. These can be found in \cite{Patton2006}.

	\subsection{Estimation strategy}
	
	The final dynamic copula realized GARCH model defines a dynamic parametric model for the joint distribution. The joint likelihood is defined as
	\begin{align}
	\mathcal{L}(\theta)   \equiv \sum_{t=1}^T\log \mathbf{f}_t(\mathbf{X_t};\theta) =& \sum_{t=1}^T \log f_{1t}(X_{1t};\theta_1)+\sum_{t=1}^T \log f_{2t}(X_{2t};\theta_2) \\
	&+\sum_{t=1}^T \log \mathbf{c}_t(F_{1t}(X_{1t};\theta_1),F_{2t}(X_{2t};\theta_2);\theta_c),
	\end{align}
	where $\theta=(\phi',\gamma')'$ is vector of all parameters to be estimated, including parameters of the marginal distributions $\phi$ and parameters of the copula, $\gamma$. The parameters are estimated using a two-step estimation procedure, generally known as multi-stage maximum likelihood (MSML) estimation, first estimating the marginal distributions and then estimating the copula model conditioning on the estimated marginal distribution parameters. While this greatly simplifies the estimation, inference on the resulting copula parameter estimates is more difficult than usual as the estimation error from the marginal distribution must be taken into account. In result, MSMLE is asymptotically less efficient than one-stage MLE, however as discussed by \cite{patton2006b}, this loss is not great in many cases. Moreover, bootstrap methodology can be used for statistical inference. 
	
\subsubsection{Semiparametric models}
	
	One of the appealing alternatives to a fully parametric model is to estimate univariate distribution non parametrically, for example by using the empirical distribution function. Combination of a nonparametric model for marginal distribution and parametric model for the copula results in a semiparametric copula model, which we use for comparison to its fully parametric counterpart. Forecasts based on a semiparametric estimation where nonparametric marginal distribution is combined with parametric copula function are not common in economic literature thus it is interesting to compare it in our modelling strategy. For the margins of the semi-parametric models, we use the non-parametric empirical distribution $F_i$ introduced by \cite{Genest:1995fk}\footnote{The asymptotic properties of this estimator can be found in \cite{Chen:2006kx}.}, which consists of modelling the marginal distributions by the (rescaled) empirical distribution.
\begin{equation}
\hat{F}_i(z)= \frac{1}{T+1}  \sum_{t=1}^T \mathbf{1} \{\hat{z}_{it} \leq z\}
\end{equation}
In this case, the parameter estimation is conducted via maximizing likelihood
\begin{equation}
	\mathcal{L}(\gamma)   \equiv \sum_{t=1}^T \log \mathbf{c}_t(\hat{U}_{1t},\hat{U}_{2t};\gamma).
	\end{equation}
As it can be seen, the likelihood reduces in estimating the copula parameters only. However, we should note that the inference on parameters is more difficult than usual, hence we rely on bootstrap inference as advocated in \cite{patton2006b}.

	\subsection{Copula selection}

An important issue when working with copulas is the selection of the best copula from the pool. Several methods and tests have been proposed for selection procedure. The methods proposed by \cite{roncalli00}  are based on distance from empirical copula. \cite{CHEN2005} propose the use of pseudo-likelihood ratio test for selecting semiparametric multivariate copula models.\footnote{Although some authors use AIC (or BIC) for choosing among two copula models, selection based on these indicators may hold only for the particular sample in consideration (due to their randomness) and not in general. Thus, proper statistical testing procedures are required [see \cite{CHEN2005}].} A test on conditional predictive ability (CPA) is proposed by \cite{Giacomini2006}. This is a robust test which allows to accommodate both, unconditional and conditional objectives. Recently, \cite{Diks:2010fk} have proposed a test for comparing predictive ability of competing copulas. The test is based on Kullback-Leibler
information criterion (KLIC) and its statistics is a special case of the unconditional version of \cite{Giacomini2006}. 

As our main aim is to use the model for forecasting, out-of-sample performance of models will be tested by CPA, which considers the forecast performance of two competing models conditional on their estimated parameters to be equal under the null hypothesis
\begin{align}
H_0&: E[ \hat{\mathbf{L}}] = 0 \\
H_{A1}&: E[ \hat{\mathbf{L}}] > 0 \mbox{ and } H_{A2}: E[ \hat{\mathbf{L}}] < 0,
\end{align}
where $\hat{\mathbf{L}} =  \log \mathbf{c}_1(\hat{\mathbf{U}}_,\hat{\gamma}_{1t}) - \log \mathbf{c}_2(\hat{\mathbf{U}}_,\hat{\gamma}_{2t})$. Other advantages of this test are the possibilities to use it for both nested and non-nested models, and also for comparison of parametric and semiparametric models. The asymptotic distribution of the test statistic is $N(0,1)$ and we compute the asymptotic variance using HAC estimates to correct for possible serial correlation and heteroskedasticity in the differences in log-likelihoods.

\section{Data description}

The data set consists of the 5-minute prices of the Prague PX and German DAX cash indices over the period January 3, 2008 until May 31, 2013, covering the recent recession. We synchronise the data using the time stamp matching, and eliminate transactions executed on Saturdays, Sundays, holidays, December 24 to 26, and December 31 to January 2 due to low activity on these days, which could lead to estimation bias. Hence, in our analysis we work with data from 1349 days. 
\begin{figure}
\begin{center}
\includegraphics[width=0.47\textwidth]{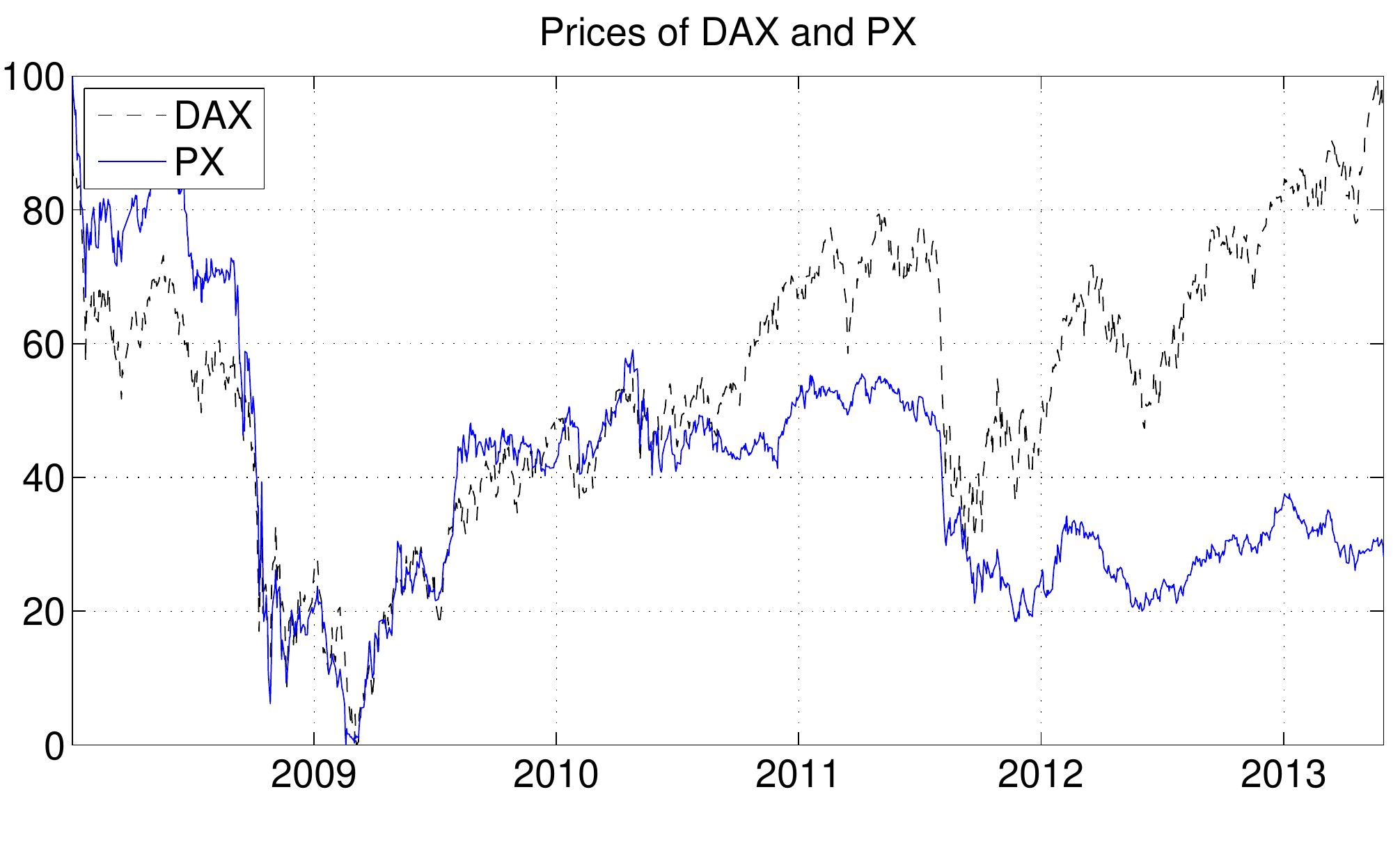}
\includegraphics[width=0.47\textwidth]{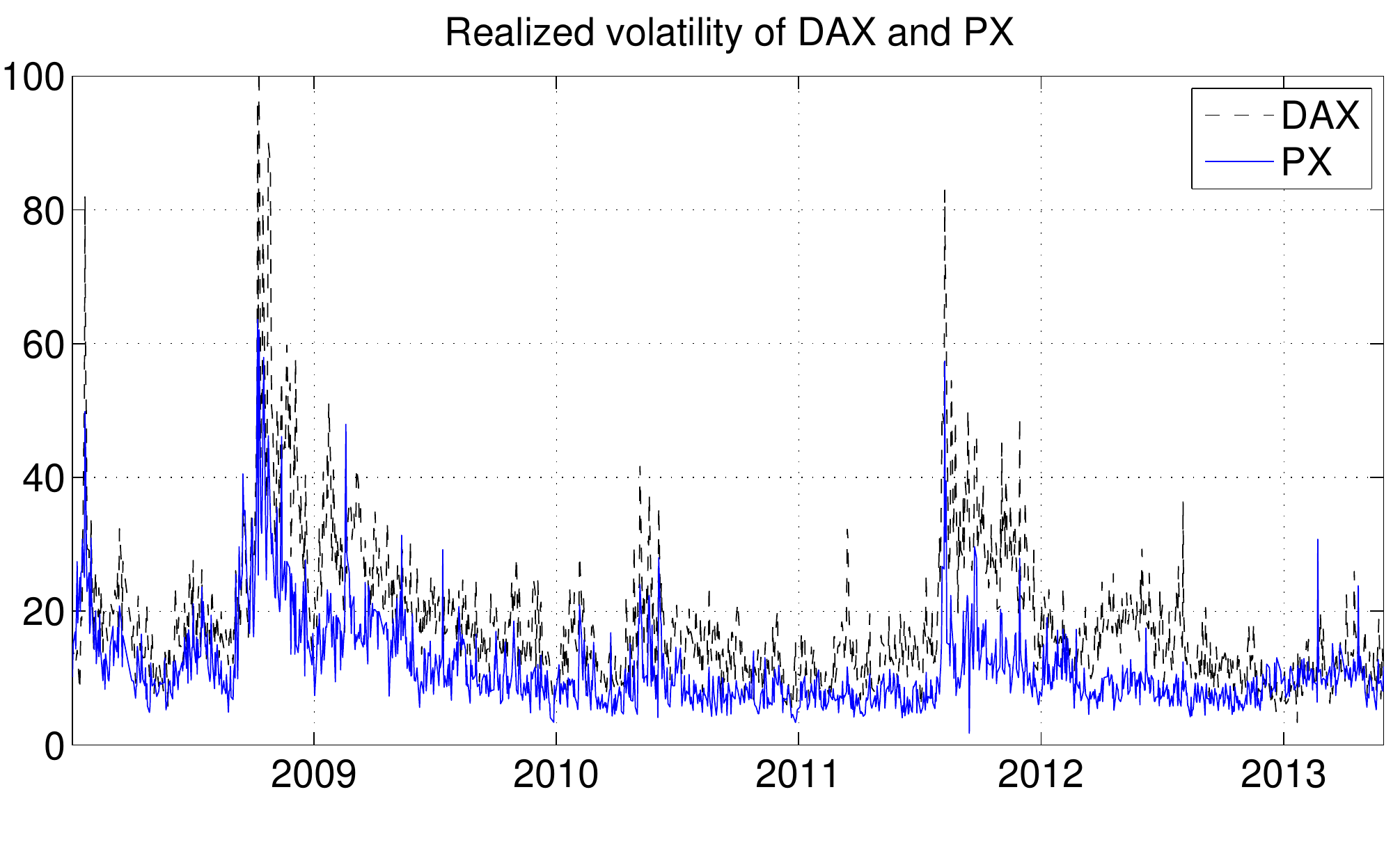}
\end{center}
\caption{Normalized prices and annualized realized volatilities of the DAX and PX over the sample period extending from January 3, 2008 until May 31, 2013.}
\label{fig:prices}
\end{figure}
For the empirical model, we need two time series, namely daily returns and realized variance to be able to estimate the realized GARCH model in margins. For this, obtain daily returns as a sum of logarithmic intraday returns, hence we work with open-close returns. Realized variance is computed as a sum of squared 5-minute intraday returns. Figure \ref{fig:prices} plots the development of prices of the PX and DAX together with its realized volatility. Note that plot of prices is normalized so we can compare the movements, and for the plot of realized volatility, we use daily volatility annualized according to the convention $100 \times \sqrt{250 \times RV_t}$. Strong time-varying nature of the volatility can be noticed immediately for both PX as well as DAX indices. Realized volatility of the DAX is larger in average when compared to the volatility of the PX index. Otherwise the volatility has similar distributional properties for both indices. 

\section{Empirical Results}

\begin{table}[h] 
\footnotesize
\begin{tabular}{lrrrrlrrrr}\toprule 
 & \multicolumn{2}{c}{DAX}& \multicolumn{2}{c}{PX}&&   \multicolumn{2}{c}{DAX}& \multicolumn{2}{c}{PX}\\ 
\midrule 
& \multicolumn{4}{c}{AR(2)}&&\multicolumn{4}{c}{AR(2)}\\
\cmidrule{2-5} \cmidrule{7-10}  
$c$ &   0.0000 &(0.08)& -0.0005 &(-0.96)&$c$  & 0.0000 &(0.08)& -0.0005 &(-0.96)\\ 
$\alpha_1$  &- &- & 0.0842 &(3.10)&$\alpha_1$ & - &- & 0.0842 &(3.10)\\ 
$\alpha_2$   &- &- & -0.1069 &(-3.94)&$\alpha_2$   &- &- & -0.1069 &(-3.94)\\ 
\cmidrule{2-5} \cmidrule{7-10} 
& \multicolumn{4}{c}{Realized GARCH(1,1)}&&\multicolumn{4}{c}{GARCH(1,1)}\\
\cmidrule{2-5} \cmidrule{7-10}  
$\omega$  & 0.2000 &(7.64)& 0.1794 &(6.82)&$\kappa$  & 0.0113 &(2.90)& 0.0124 &(3.28)\\ 
$\beta$  & 0.5746 &(21.19)& 0.6600 &(23.31)&$\phi$ & 0.0852 &(6.15)& 0.1498 &(7.43)\\ 
$\gamma$  & 0.4072 &(14.22)& 0.3399 &(11.55)&$\psi$ & 0.9022 &(61.01)& 0.8424 &(45.39)\\  
$\xi$  & -0.5376 &(-12.05)& -0.5834 &(-12.13)&&-&-&-&-\\ 
$\phi$  & 0.9655 &(22.75)& 0.8996 &(23.46)& &-&-&-&-\\ 
$\tau_1$  & -0.1691 &(-13.26)& -0.1414 &(-9.44)& &-&-&-&-\\  
$\tau_2$  & 0.0717 &(8.97)& 0.0943 &(10.23)& &-&-&-&-\\ 
$\nu$  & 13.6919 &(3.09)& 7.3569 &(5.27)& &-&-&-&-\\  
$\lambda$  & -0.1161 &(-3.39)& -0.0830 &(-2.30)& &-&-&-&-\\ 
\cmidrule{2-5} \cmidrule{7-10}  
$\mathcal{LL}_{r,x}$  &\multicolumn{2}{c}{    -2461.00 } &\multicolumn{2}{c}{    -2545.41 } & &\multicolumn{2}{c}{    - }& \multicolumn{2}{c}{    - }\\ 
$\mathcal{LL}_{r}$  &\multicolumn{2}{c}{  -1606.75 } &\multicolumn{2}{c}{    -1508.11 } &$\mathcal{LL}$  &\multicolumn{2}{c}{    -1682.13 } &\multicolumn{2}{c}{    -1522.45 } \\ 
$AIC_r$  &\multicolumn{2}{c}{     3231.50 } &\multicolumn{2}{c}{     3034.23 }&$AIC$   &\multicolumn{2}{c}{     3370.27 } &\multicolumn{2}{c}{     3050.90 } \\ 
$BIC_r$   &\multicolumn{2}{c}{    3278.37 } &\multicolumn{2}{c}{   3081.09 } &$BIC$   &\multicolumn{2}{c}{     3385.89 } &\multicolumn{2}{c}{     3066.52 } \\ 
  \bottomrule  
\end{tabular}   
\caption{Parameter estimates from AR($2$) \emph{log-linear} Realized GARCH(1,1) and benchmark GARCH(1,1) with innovations distributed \emph{skew-t} and normal respectively. \emph{t}-statistics reported in parentheses.}  
\label{tab:condVarModel}  
\end{table} 
Before modeling the dependence structure between the PX and DAX, we need to model their conditional marginal distributions first. Considering general AR models up to five lags, we find AR(2) to best capture the time-varying dependence in mean of PX, while DAX has a constant mean. These results are in line with previous research \citep{barunik2008neural}. Table \ref{tab:condVarModel} summarizes the Realized-GARCH(1,1) fit for both PX and DAX. In addition, benchmark GARCH(1,1) model is fit to the data for comparison. All the estimated parameters are significantly different from zero. By observing partial log-likelihood $\mathcal{LL}_{r}$ as well as information criteria, we can see that including realized measures into the GARCH model improves the fits significantly. This is crucial for copulas, as we need to specify the best possible model in the margins to make sure there is no univariate dependence left. If a misspecified model is used for the marginal distributions, then the probability integral transforms will not be uniformly distributed and this will result in copula misspecification. For the estimated standardized residuals from the AR(2) realized GARCH(1,1), we consider both parametric and nonparametric distributions as motivated earlier in the text. 

 \subsection{Time-varying dependence between DAX and PX}
 
 Before specifying a functional form for time-varying copula function, we test for the presence of time-varying dependence using the simple approach based on the ARCH LM test. The test statistics is computed from the OLS estimate of the covariance matrix and critical values are obtained using \textit{i.i.d.} bootstrap (for detailed information, consult \cite{Patton2012}). Computing the test for the time-varying dependence between the DAX and PX up to $p=10$ lags, we find the joint significance of all coefficients. Thus we can conclude that there is evidence against constant correlation for the DAX and PX.
\begin{table}
\footnotesize
\begin{center} 
\begin{tabular*}{\textwidth}{@{\extracolsep{\fill}}llcccccccc}
\toprule 
& &  \multicolumn{3}{c}{Parametric} & & \multicolumn{3}{c}{Semiparametric}\\ 
\cmidrule{2-5} \cmidrule{7-10}  
		& & & & \\
 \multicolumn{4}{l}{\bf{Constant copula}} \\ 
 	& & \multicolumn{2}{c}{Est. Param} & $\mathcal{\log L}$ & & \multicolumn{2}{c}{Est. Param} & $\mathcal{\log L}$ & \\
	\cmidrule{3-5} \cmidrule{7-10} 
Normal & 		$\rho$ &  0.6042 & (0.0188) & \textbf{305.90 } & &   0.6053 & (0.0157) & \textbf{       307.83 } \\ 
		\cmidrule{3-5} \cmidrule{7-10} 
Clayton  &$\kappa$ &  0.8596 &(0.0591) & \textbf{       221.98 }&  &0.9258 &(0.0560) & \textbf{       232.08 }\\ 
		\cmidrule{3-5} \cmidrule{7-10} 
RGumb & $\kappa$ &  1.5819 &(0.0385) & \textbf{       265.24 }  & &  1.6130 &(0.0323) &  \textbf{       272.99 }\\ 
		\cmidrule{3-5} \cmidrule{7-10} 
Student's $t$ & $\rho$ &  0.5960 &(0.0192) &  &&0.6076 &(0.0139) & \\ 
		   & $\nu^{-1}$ &  0.0100 &(0.0224) &   \textbf{       305.53 } &&0.0100 &(0.0042) & \textbf{       307.43 } \\ 
		\cmidrule{3-5} \cmidrule{7-10} 
Sym. Joe-Clayton		& $\tau^L$ &  0.3667 &(0.0295)&  && 0.3991 &(0.0318)\\ 
		& $\tau^U$ &  0.3514 &(0.0366)&  \textbf{       273.96 } && 0.3552&(0.0356) &  \textbf{       279.76 } \\ 	
		& & & & \\
 \multicolumn{4}{l}{\bf{``GAS" time-varying copula}} \\ 
  	& & \multicolumn{2}{c}{Est. Param} & $\mathcal{\log L}$ & & \multicolumn{2}{c}{Est. Param} & $\mathcal{\log L}$ & \\
	\cmidrule{3-5} \cmidrule{7-10} 
$RGumb_{GAS}$ 	& $\hat{\omega}$ &  -0.0466 &  (0.1245) & & &   -0.0037 & (0.1103) & \\ 
				&$\hat{\alpha}$ &  0.0466 & (0.0520) &  & & 0.0207 & (0.0496) & \\ 
 				&$\hat{\beta}$ &  0.9139 & (0.2137) & \textbf{266.69 } & &  0.9927 & (0.2491) & \textbf{275.95 }\\ 
\cmidrule{3-5} \cmidrule{7-10} 
$N_{GAS}$	&$\hat{\omega}$ &  0.0121 & (0.2883) && &  0.0131 & (0.3311) &  \\ 
				&$\hat{\alpha}$ &  0.0244 & (0.0371) & & &  0.0274 & (0.0377) & \\ 
				&$\hat{\beta}$ &  0.9911 & (0.2064) &  \textbf{      312.28 } & &  0.9907 & (0.2198) &  \textbf{      314.54 }\\ 
\cmidrule{3-5} \cmidrule{7-10} 
$t_{GAS}$	&$\hat{\omega}$ &  0.1466 & (0.2740) & & &  0.1159 & (0.2481) &  \\ 
				&$\hat{\alpha}$ &  0.0662 & (0.0379) & & &  0.0755 & (0.0375) & \\ 
				&$\hat{\beta}$ &  0.8936 & (0.1941) & & &  0.9188 & (0.1720) &  \\ 
				&$\hat\nu^{-1}$ &  0.0115 & (0.0069) & \textbf{      311.41 }& &  0.0120 & (0.0092) &  \textbf{      313.43 } \\ 
  \bottomrule  
\end{tabular*}   
\end{center} 
\caption{ Constant and time-varying copula model parameter estimates with AR($2$)-Realized GARCH(1,1) model for both fully parametric and semiparametric cases. Bootstrapped standard errors are reported in parentheses.}  
\label{tab:UnivRGTvDAX-PX}  
\end{table} 
Motivated by this finding, we estimate three time-varying copula functions, namely Normal, rotated Gumbel and Student's $t$ using the GAS framework described in the methodology part. As a benchmark, we also estimate the constant copulas to be able to compare the time-varying models against the constant ones. While semiparametric approach is empirically interesting and not often used in literature, we use it for all the estimated models as well.

Table \ref{tab:UnivRGTvDAX-PX} shows the fit from all estimated models. Starting with constant copulas, all the parameters are significantly different from zero and Normal copula seems to describe the DAX and PX indices best according to the highest log-likelihood. Semiparametric specifications combining nonparametric distribution in margins with parametric copula function bring further improvement in the log-likelihoods. Importantly, time-varying specifications bring large improvement in log-likelihoods and confirm strong time-varying dependence between the DAX and PX indices. Due to the large number of degrees of freedoms, $t_{GAS}$ copula in fact converges to the normal one $N_{GAS}$, and thus time-varying normal copula again best describes the data. 

This is interesting finding, as it confirms that after proper models for the dependence in margins of the distribution, there is no asymmetry left and the PX-DAX bivariate distribution is standard normal. To study the goodness of fit for all the specified models, we use\footnote{The results of the in-sample goodness of fit tests are available upon request from authors. We do not include them in text to save the space.} Kolmogorov-Smirnov (KS) and Cramer-von Mises (CvM) test statistics with $p$-values obtained from 1000 simulations. None of the fully parametric models is rejected, while most of the semiparametric models are rejected with exception of constant student's $t$, Sym. Joe-Clayton and time-varying student's $t$. These results suggest that fully parametric models with realized GARCH and parametric distribution in margins are all well-specified. Thus realized GARCH seems to very well model all the dependence in margins, which is crucial for the good specification of the model in the copula-based approach. Semiparametric models are interestingly rejected and are not specified well, except for few mentioned cases. This is in line with results of \cite{Patton2012}, who finds rejections in semiparametric specifications on the U.S. indices data. Still, both tests strongly support the realized GARCH time-varying GAS copulas for modeling the joint distribution between DAX and PX.

\begin{figure}[h]
\begin{center}
\includegraphics[width=0.5\textwidth]{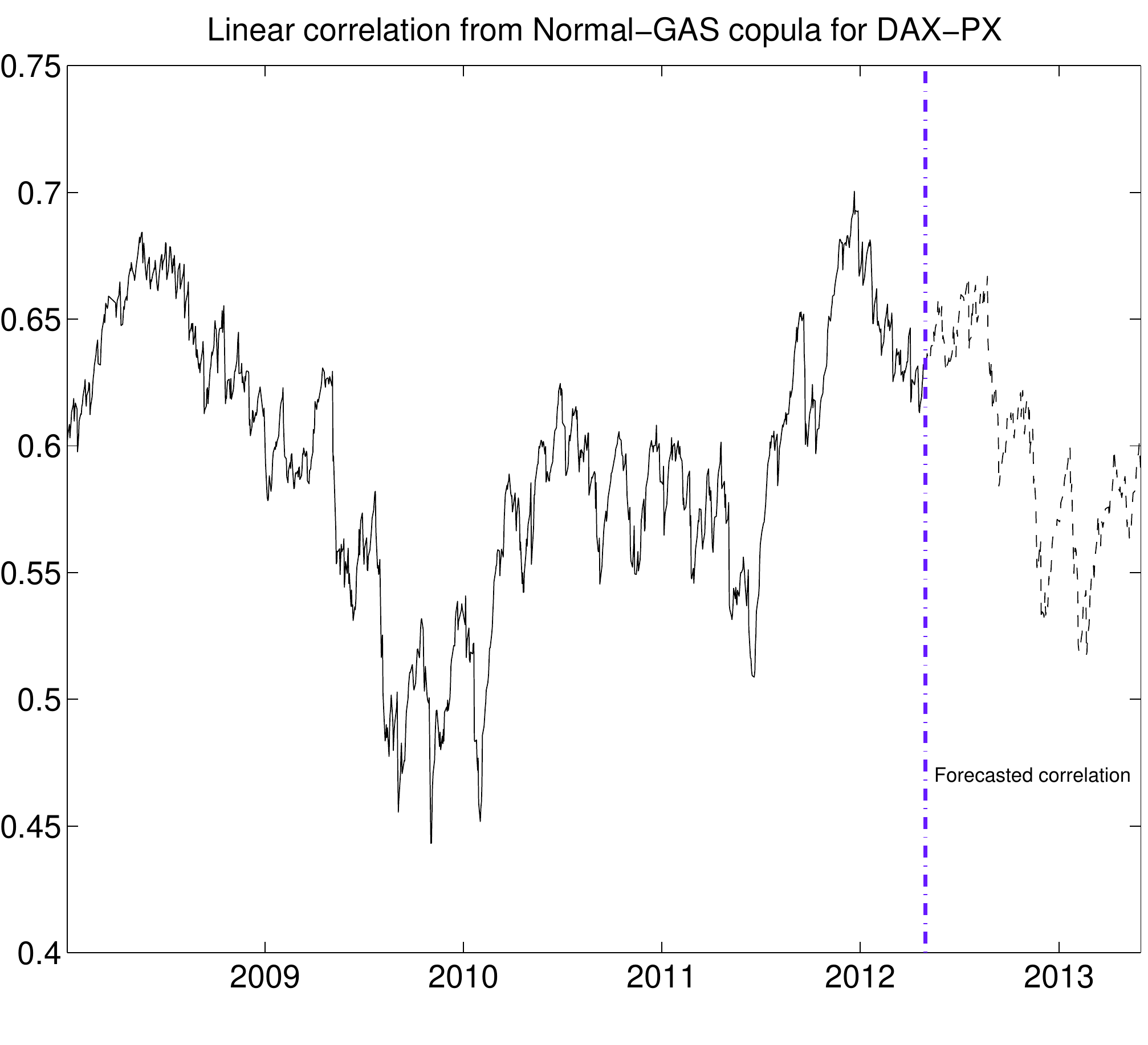}
\end{center}
\caption{ Linear correlation from time-varying Normal GAS copula. The vertical dashed line separates the in-sample from the out-of-sample (forecasted) part. }
\label{fig:correlations}
\end{figure}

\subsection{Out-of-sample comparison of the proposed models}

While it is important to have a well-specified model which describes the data, most of the times we are interested in using the model for forecasting. Thus we conduct an out-of-sample evaluation of the proposed models. For this, the sample is divided into two periods. The first, in-sample period, is used to obtain parameter estimates from all models and spans from January 3, 2008 to May 2, 2012. The second, out-of-sample period is then used for evaluation of forecasts. Due to highly computationally intensive estimation of the models, we restrict ourselves to a fixed window evaluation, where the models are estimated only once, and all the forecasts are done using the recovered parameters from this fixed in-sample period. This makes it even harder for the models to perform well in the highly dynamic data.

For the out-of-sample forecast evaluation\footnote{The results of the out-of-sample forecast evaluation are available upon request from authors. We do not include them in text to save the space, and no to distract the reader from the main results.}, we use the conditional predictive ability (CPA) test of \cite{Giacomini2006}. The time-varying copula models outperforms significantly the constant copula models in out-of-sample evaluation. This holds both for parametric and semiparametric cases. Thus time-varying copulas have much stronger support for forecasting the dynamic distribution of the DAX and PX. When comparing the different time-varying copula functions, the test is not so conclusive. While student's $t$ and normal statistically outperform Rotated gumbel, the forecasts from student's $t$ can not be statistically distinguished from the normal copula. Time-varying normal and student's $t$ copulas are thus best in the forecasting exercise. Finally, forecasts from parametric models statistically outperform those from semiparametric ones.

Thus, the out-of-sample results confirm the in-sample ones, which is a good sign of proper model fit. The joint distribution of the PX and DAX indices is best modeled with the AR(2)-realized GARCH(1,1) time-varying normal copula model.

Having correctly specified the empirical model capturing the dynamic joint distribution between the DAX and PX, we can proceed to studying the pair. Figure \ref{fig:correlations} plots the time-varying correlations implied by our model with normal GAS copula. The dependence is generally strong, and also has strong time-varying nature during the studied period. During the last quarter of the year 2008, when stock markets were declining due to the Lehman Brother's crash, the correlation of the PX and DAX markets rose nearly to 0.7. In the following year, it dropped to 0.45 levels and from the year 2010 rose back to 0.7 again. 

This result has serious implications for investors as it suggests that diversification possibilities are rapidly changing over past few years during the financial crisis. We are going to utilize the results and study the possible economic benefits of the modeling strategy. 

\section{Economic implications: Time-varying diversification benefits and VaR}

While it is important to have statistically correct fits, or even good out-of-sample forecasts, the crucial question is whether it translates to economic benefits. Here we test our proposed methodology for economic implications. First, we quantify the risk of an equally weighted portfolio composed from the DAX and PX, and second, we study the benefits from diversification to see how the strongly varying correlation affect them. This is mainly interesting to the international investors considering the Czech PX stock market index and the German DAX index in the portfolio.

\subsection{Quantile forecasts}

Quantile forecasts are central to risk management decisions due to a widespread Value at risk (VaR) measurement. VaR is defined as the maximum expected loss which may be incurred by a portfolio over some horizon with a given probability. Let $q_t^{\alpha}$ denote an $\alpha$ quantile of a distribution. VaR of a given portfolio at time $t$ is then simply 
\begin{equation}
q_t^{\alpha} \equiv  F_t^{-1} ({\alpha}), \text{for } {\alpha} \in (0,1).
\end{equation}

\begin{figure}
\begin{center}
\includegraphics[width=0.4\textwidth]{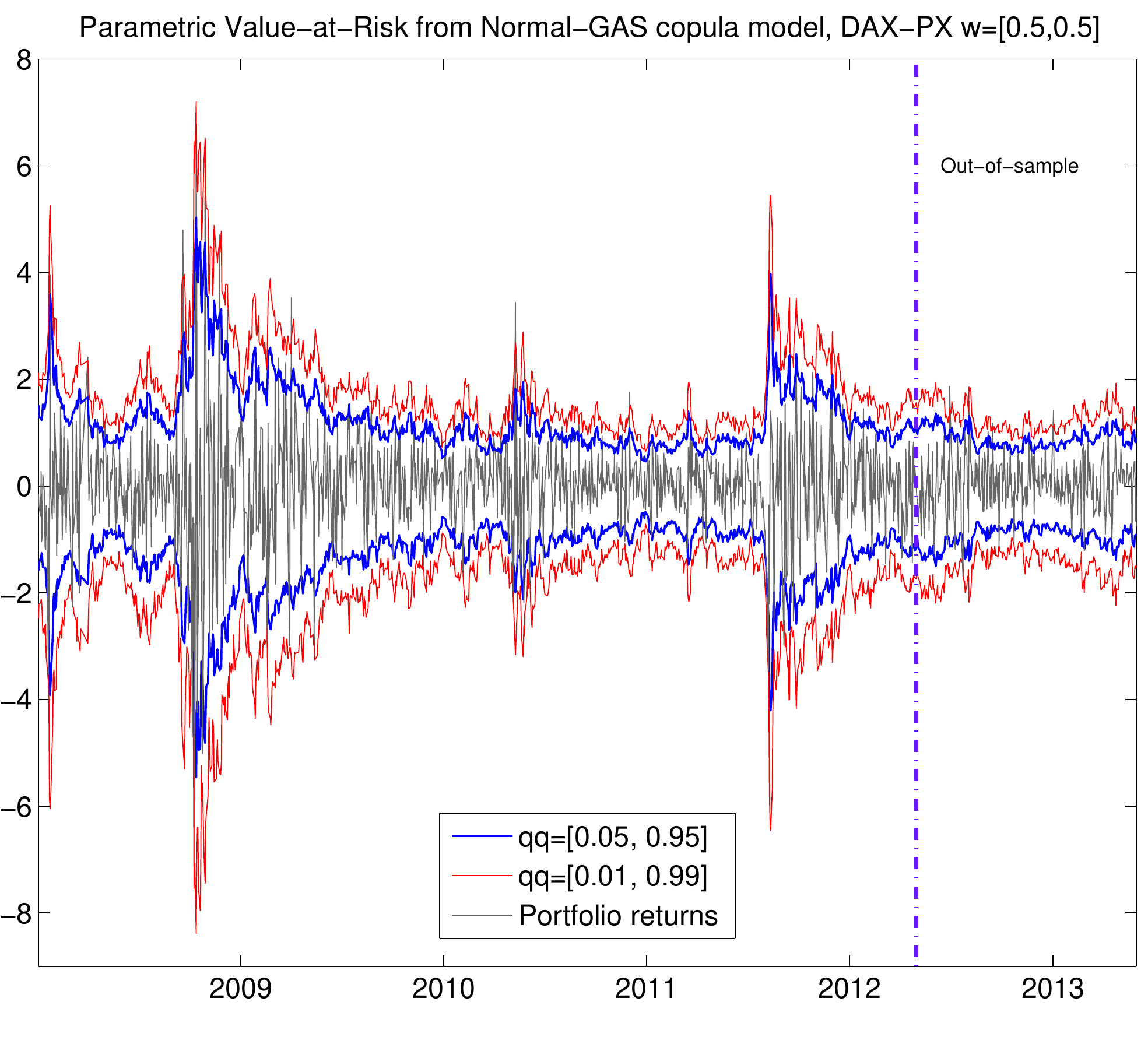}
\includegraphics[width=0.4\textwidth,height=5.59cm]{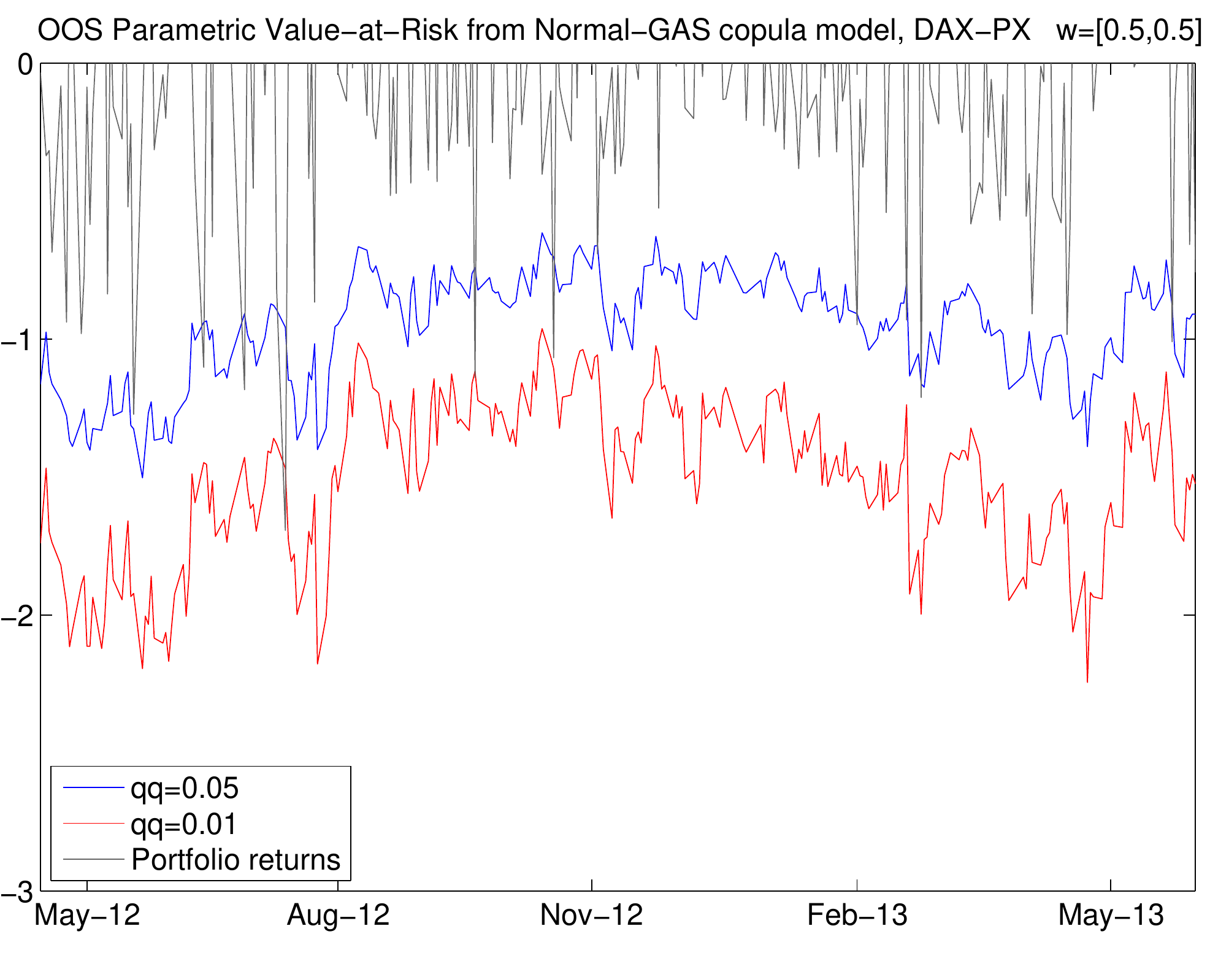}
\end{center}
\caption{Value-at-Risk implied by Realized-GARCH time-varying Normal GAS copulas. The portfolio consists of an equal-weight amount of DAX and PX and the estimation is made for quantiles 1\% and 5\%.}
\label{fig:VaRs}
\end{figure}
Thus choice of the distribution is crucial to VaR calculation. For example assuming normal distribution may lead to underestimation of the VaR. Our objective is to estimate one-day-ahead VaR of an equally weighted portfolio composed from the DAX and PX returns as $Y_t = 0.5 X_{1t} + 0.5 X_{2t}$, which has conditional time-varying joint distribution $F_t$. In the previous analysis, we have found that the realized GARCH model with time-varying normal GAS copula well fits and forecasts the data, thus we use it in VaR forecasts to see whether it correctly forecasts also the joint distribution. As there is no analytical formula, which can be used for this, we rely on Monte Carlo approach, where we simply simulate the future conditional joint distribution from the estimated models.

\begin{table}[t!]
\footnotesize
\caption{Out-of-sample VaR evaluation. Empirical quantile $\hat{C}_{\alpha}$, estimated Giacomini and Komunjer (2005) $\hat{L}$, logit DQ statistics and its 1000$\times$ simulated $p$-val are reported. $\hat{L}$ is moreover tested with Diebold-Mariano statistics with Newey-West estimator for variance. All models are compared to $N_{GAS}$, while models with significantly less accurate forecasts at 95\% level are reported in bold.}
\centering
\resizebox{0.98\textwidth}{!}{
\begin{tabular*}{1.2\textwidth}{@{\extracolsep{\fill}}lrrrrrrrrrrrrrr}
\toprule 
  &\multicolumn{6}{c}{Parametric} & & \multicolumn{6}{c}{Semiparametric}  \\ 
\cmidrule{2-7}\cmidrule{9-14}
 & 1\% & 5\% & 10\% & 90\% & 95\% & 99\% & & 1\% & 5\% & 10\% & 90\% & 95\% & 99\%\\  
\cmidrule{2-7}\cmidrule{9-14}
\multicolumn{3}{l}{$\mathbf{RGumb_{GAS}}$} \\
 $\hat{C}_{\alpha}$ &0.004 & 0.033 & 0.074 & 0.915 & 0.948 & 0.989 & & 0.004 & 0.033 & 0.070 & 0.922 & 0.948 & 0.989 \\
 $\hat{L}$ &\textbf{0.017} & 0.058 & 0.095 & 0.089 & 0.054 & 0.016 & & \textbf{0.017} & 0.058 & 0.094 & 0.089 & 0.054 & 0.016 \\
 DQ &1.301 & 5.023 & 5.941 & 13.362 & 6.764 & 0.396 & & 1.301 & 5.023 & 11.758 & 12.432 & 6.764 & 0.396 \\
 $p$-val &0.972 & 0.541 & 0.430 & 0.038 & 0.343 & 0.999 & & 0.972 & 0.541 & 0.068 & 0.053 & 0.343 & 0.999 \\
&\\
\multicolumn{3}{l}{$\mathbf{t_{GAS}}$} \\
 $\hat{C}_{\alpha}$ &0.004 & 0.037 & 0.078 & 0.919 & 0.948 & 0.989 & & 0.004 & 0.037 & 0.074 & 0.911 & 0.952 & 0.989 \\
 $\hat{L}$ &0.017 & 0.058 & 0.094 & 0.088 & 0.054 & 0.016 & & \textbf{0.017} & 0.057 & 0.094 & 0.089 & 0.054 & 0.016 \\
 DQ &1.301 & 4.663 & 5.781 & 12.735 & 6.764 & 0.396 & & 1.301 & 4.663 & 5.941 & 14.299 & 5.779 & 0.396 \\
 $p$-val &0.972 & 0.588 & 0.448 & 0.047 & 0.343 & 0.999 & & 0.972 & 0.588 & 0.430 & 0.026 & 0.448 & 0.999 \\
&\\
\multicolumn{3}{l}{$\mathbf{N_{GAS}}$} \\
 $\hat{C}_{\alpha}$&0.007 & 0.037 & 0.070 & 0.922 & 0.948 & 0.989 & & 0.011 & 0.033 & 0.074 & 0.926 & 0.952 & 0.989 \\
 $\hat{L}$ &0.016 & 0.058 & 0.094 & 0.089 & 0.054 & 0.016 & & 0.016 & 0.058 & 0.095 & 0.089 & 0.054 & 0.016 \\
 DQ &0.330 & 4.663 & 11.758 & 12.432 & 6.764 & 0.396 & & 0.396 & 5.023 & 5.941 & 11.655 & 5.779 & 0.396 \\
 $p$-val &0.999 & 0.588 & 0.068 & 0.053 & 0.343 & 0.999 & & 0.999 & 0.541 & 0.430 & 0.070 & 0.448 & 0.999 \\
 \bottomrule
\end{tabular*}
}
\label{tab1VaR}
\end{table}

While quantile forecasts can be readily evaluated by comparing their actual (estimated) coverage 
$\hat{C}_{\alpha} = 1/n \sum_{n=1}^T 1(y_{t,t+1}<\hat{q}^{\alpha}_{t,t+1})$, against their nominal coverages rate, 
$C_{\alpha} = E[1(y_{t,t+1}<q^{\alpha}_{t,t+1})]$, this approach is unconditional and does not capture the possible dependence in the coverage rates. Number of approaches has been proposed for testing the appropriateness of quantiles conditionally, for the best discussion see \cite{berkowitz2011evaluating}. In our out-of-sample VaR testing, we use an approach originally proposed by \cite{engle2004caviar}, who use the $n$-th order autoregression $I_{t} = \omega + \sum_{k=1}^n \beta_{1k}I_{t-k} + \sum_{k=1}^n \beta_{2k} q^{\alpha}_{t-k+1} + u_t$, 
where $I_{t+1}$ is $1$ if $y_{t+1}<q_t^{\alpha}$ and zero otherwise. While hit sequence $I_t$ is a binary sequence, $u_t$ is assumed to follow a logistic distribution and we can estimate it as a simple logit model and test whether $Pr(I_t = 1) = q_t^{\alpha}$. To obtain the $p$-values, we rely on simulations as suggested by \cite{berkowitz2011evaluating} and we refer to this test as a $DQ$ test in the results. 

Moreover, we evaluate the accuracy of VaR forecasts statistically by defining the expected loss of VaR forecast made by a forecaster $m$ as
\begin{equation}
L_{\alpha,m} = E\left[ \alpha - 1\left( y_{t,t+1}<q^{\alpha,m}_{t,t+1} \right) \right] \left[ y_{t,t+1}-q^{\alpha,m}_{t,t+1} \right],
\end{equation}
which has been proposed by \cite{giacomini2005evaluation}. Then, differences in the values of $L_{\alpha,m}$ can be tested using \cite{diebold2002comparing} approach, where we test the null hypothesis that the loss function of a benchmark forecaster is the same as the loss function of the tested forecaster $m$, under the alternative that benchmark forecaster is more accurate than the competing one. 

Table \ref{tab1VaR} reports out-of-sample VaR evaluation of all models, and Figure \ref{fig:VaRs} illustrates the 1\% and 5\% estimated quantiles of the portfolio. We can see that all the time-varying models are well specified and the conditional quantile forecasts from them are not rejected by the DQ test. For the statistical testing, we use time-varying normal copula as a benchmark forecaster and test all the other models against it. When looking at the loss functions $\hat{L}_{\alpha,m}$, we can see that all the quantiles implied by the different models can not be distinguished from each other statistically, except the 1\% quantile. This is mainly because student's $t$ copula has large number of degrees of freedom basically converged to the normal one. Thus overall, AR(2)-realized GARCH(1,1) with time-varying copula models are able to describe and forecast the quantiles of the PX-DAX distribution very well. 

\subsection{Time varying diversification benefits}

In case the dependence of the assets is changing over time strongly, it needs to translate to changing diversification benefits as well. Unlike VaR, expected shortfall satisfies the sub-additivity property and is a coherent measure of risk. Motivated by these properties, \cite{christoffersen2012potential} propose a measure capturing the dynamics in diversification benefits based on expected shortfall. The conditional diversification benefit (CDB) for a given probability level $\alpha$ is defined by
\begin{figure}[h]
\begin{center}
\includegraphics[width=0.7\textwidth]{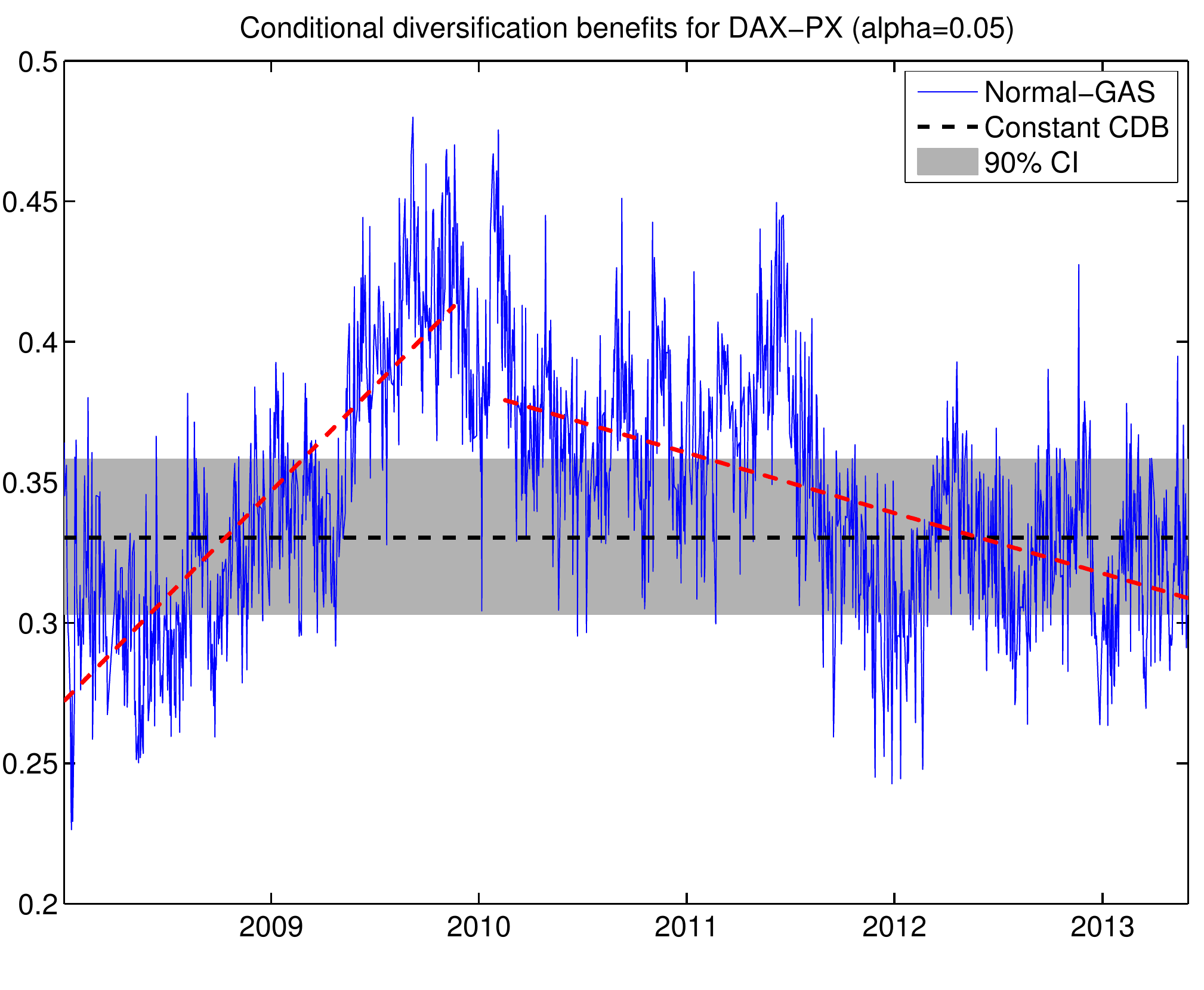}
\end{center}
\caption{Conditional diversification benefits, $CDB_t^{0.05}$ using time-varying normal copula together with bootstraped confidence band for the constant conditional diversification benefits.}
\label{fig:cdb}
\end{figure}
\begin{equation}
CDB_t^{\alpha} = \frac{\overline{ES}_t^{\alpha}-ES_t^{\alpha}}{\overline{ES}_t^{\alpha}-\underline{ES}_t^{\alpha}},
\end{equation}
where $ES_t^{\alpha}$ is expected shortfall of the portfolio at hand,
\begin{equation}
ES_t^{\alpha} \equiv E[Y_t|\mathcal{F}_{t-1}, Y_t \leq F_t^{-1} ({\alpha})], \text{for } \alpha \in (0,1),
\end{equation}
$\overline{ES}_t^{\alpha}$ is upper bound of the portfolio expected shortfall being the weighted average of the asset's individual expected shortfalls, and $\underline{ES}_t^{\alpha}$ lower bound on the expected shortfall being the inverse cumulative distribution function for the portfolio. In other words, this lower bound corresponds to the case where the portfolio never loses more than its $\alpha$ distribution quantile. The measure is designed to stay within $[0,1]$ interval, and is increasing in the level of diversification benefits. When the CDB is equal to zero, there are literarily no benefits from diversification, when it equals one, the benefits from diversification are highest possible. 

Figure \ref{fig:cdb} plots the conditional diversification benefits for the PX and DAX portfolio implied by our empirical model for $\alpha=5\%$. Similarly to the VaR case, as there is no closed form solution to our empirical model available, we rely on the simulations for CDB computation. Encouraged by the previous results, we compute the CDB for the AR(2) realized GARCH with time-varying normal copula model. Analysis could be taken step ahead by optimizing portfolio weights for the highest diversification benefits. This is done in \cite{christoffersen2012potential}, who basically find very small increase, implying that equally weighted portfolio is usually very close to optimal if CDB is used. Also please note that here we do not exploit the full potential of dynamic asset allocation.

Diversification benefits vary over time greatly. From the beginning of the sample, the benefits from diversification between the DAX and the PX index are rising gradually until the end of the year 2009, when they start to decline. The lowest values are at the beginning of the year 2012, while from this point until the 2013, the benefits stay more or less lower.

To support our results, we also report 90\% bootstrapped confidence bands computed around a constant level of diversification benefits. Assuming the returns data are independently distributed over time with the same unconditional correlation as the PX and DAX pair, bootstrap confidence level can be conveniently computed via simulations. We use 10.000 simulations, and report the mean value together with distribution of constant conditional benefits in Figure \ref{fig:cdb}. We can see, that the time-varying nature of the conditional diversification benefits is statistically significant, as it departs from the simulated constant distribution. 

Thus contrary to the general expectation of no diversification benefits for investors considering the Czech PX index as a diversification tool for the German DAX due to very high correlation between these two stocks, we find actual benefits which are varying strongly in time.

\section{Conclusions}

This work revisits the Czech PX and German DAX stock markets dependence with the aim to study the opportunities of these two assets in portfolio management. Using an empirical model utilizing high frequency data in the time varying copulas, we study the joint conditional distribution of the PX and DAX returns. 

The final AR(2) realized GARCH(1,1) with time-varying normal copula is able to capture the dynamics accurately, yielding precise quantile forecasts. Using the crisis data, we study the time-varying correlations between the PX and DAX returns. More important, we study how the time-varying dependence translates to the conditional diversification benefits. The main result is that the possible diversification benefits are strongly varying over time, and hence even after the 2008 financial crisis, it may be economically interesting to use the DAX and PX returns for the risk diversification. This is important finding, as it is contrary to the belief that crisis caused reduction in international diversification benefits. Czech and German economies are strongly tightened as well, so one would expect that especially after the inclusion of the Czech Republic into the euro area, diversification benefits will disappear.

{\footnotesize{
\setlength{\bibsep}{3pt}
\bibliographystyle{chicago}
\bibliography{Bibliography}
}}

\end{document}